\documentclass[aps,pra,twocolumn,showpacs,noeprint]{revtex4-2}

\usepackage{graphicx,amsfonts,times,bm,amsmath,amssymb,verbatim,ifthen,braket,color,array,natbib,bm,soul}
\usepackage{array}
\usepackage{delarray}
\usepackage{subfigure}
\usepackage{xcolor} 

\usepackage{hyperref}

\begin{document}

\title{
Scaling of Computational Order Parameters in Rydberg Atom Graph States
}

\author{Zhangjie Qin}
\author{V. W. Scarola}
\email{scarola@vt.edu}
\affiliation{Department of Physics, Virginia Tech, Blacksburg, Virginia 24061, USA}

\begin{abstract}
Graph states are computationally powerful quantum states with many applications including use as resource states for measurement-based quantum computing (MBQC).  We demonstrate construction of graph states on a Rydberg atom quantum analogue simulator.  We show how an always-on interaction can be used to simultaneously entangle all Rydberg atoms into a graph state.  We construct and implement many-body computational order parameters for graph states using non-local measurement-based logic operations in the Clifford group.  The order parameters measure the efficacy of entanglement to allow MBQC on graph states of any size.  We parameterize finite-size scaling of these order parameters.  Our results define a route to efficiently test computational power in quantum devices.
\end{abstract}

\maketitle
\section{Introduction} 
\label{sec_introduciton}

Graph states are useful quantum many-body spin wavefunctions that are straightforward to construct \cite{Hein2006a}.  Consider a collection of spins defining graph vertices such that each spin is oriented along the positive-$x$ direction.  Application of two-spin entangling operations, e.g., a controlled-Z (CZ) operation between spins, establishes the graph edges.  The resulting many-body states have applications in metrology \cite{Gachechiladze2016,Shettell2020}, they help speedup variational solvers \cite{Anders2006}, they possess  \cite{Son2011,Else2012,Miller2015,Devakul2018,Stephen2019,Raussendorf2019,Daniel2020,Azses2020,RAUSSENDORF2023} symmetry protected topological order \cite{Senthil2015}, and they are useful in characterizing topological stabilizer quantum error correction codes \cite{Liao2021,Lio2022}.  Graph states are also measurement-based quantum computing (MBQC) resource states \cite{Raussendorf2001b,Raussendorf2003,Briegel2009a}. 

Starting with a graph state, MBQC \cite{Raussendorf2001b,Raussendorf2003} can proceed by preparing input on one side of the graph.  Local projective measurements throughout the graph then implement quantum logic operations on the input state.  The output is obtained on the opposing side of the graph.  It was shown that properly designed sequences of measurements \cite{Raussendorf2001b,Raussendorf2003} on graph states are equivalent to circuit-based quantum algorithms, e.g., quantum simulation subroutines \cite{Lee2022}. 

Fault tolerant MBQC on a two dimensional cluster state, an example of a graph state with translational invariance, is possible if the probability of an error in resource state preparation and measurement is kept below the threshold \cite{Raussendorf2007b}:
 \begin{align}
 \varepsilon_{\text{T}}=0.0075. 
 \label{eq_error_threshold}
 \end{align}
 The implementation of MBQC error correction discussed in Ref.~\onlinecite{Raussendorf2007b} requires large cluster states.  Experimental methods to efficiently create and characterize large graph states are therefore critical to implementing MBQC.

Characterization of large scale quantum states such as graph states remains challenging.  Conventional order parameters used in solids, e.g., magnetization, are insufficient.
Our central aim is to define practical order parameters from the perspective of information theory.  We define computational order parameters as the fidelities to implement measurement-based quantum logic operations in the Clifford group.  The order parameters define the efficacy to compute with the graph state.  The MBQC Clifford operations rely on $n$ measurements on $N$ vertices of a graph state.  They can be made global such that $n=N$.  Finite-size scaling of these order parameters establishes the domain size over which graph states can effectively be used in MBQC provided we define order parameter thresholds.  These order parameters can be used to characterize $N$-atom graph states with $n=N$ global measurements such that their scaling offers a benchmarking tool.

We consider benchmarking of graph states of neutral atoms in optical tweezer arrays as an application because these sytsems \cite{Saffman2010,BROWAEYS2020a} are rapidly scaling up in size \cite{MANETSCH2024}.  
Recent experiments \cite{BLUVSTEIN2022a} constructed and benchmarked cluster states with atoms in an optical tweezer array operating in a digital mode \cite{BROWAEYS2020a}.  Here pseudospin states were encoded into hyperfine states of Rydberg atoms.  Rydberg atoms interact via the van der Waals interaction at long distances \cite{Saffman2010,BROWAEYS2020a}.  To avoid crosstalk induced by long-range tails of the interaction, the atoms were coherently shuttled far from other atoms to entangle them pairwise into a cluster state.  The 12-atom cluster state chain was benchmarked by measuring a 3-atom stabilizer expectation value \cite{BLUVSTEIN2022a}.  

\begin{figure}[t]
\begin{center}
\includegraphics[width=0.44\textwidth]{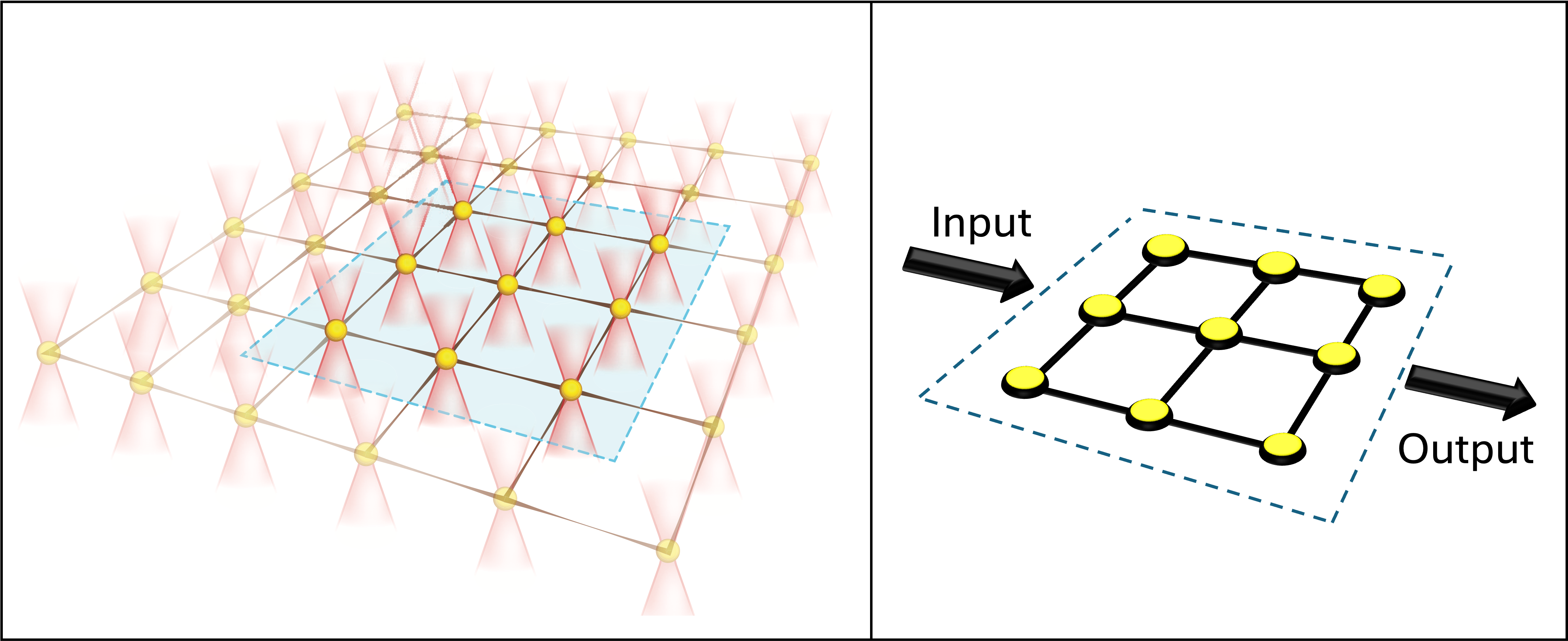}
\end{center}
\caption{Left: Schematic of atoms trapped in optical tweezers in an example pattern, a square lattice array.  Pairs of atomic hyperfine states within each atom define pseudospins.  The lines connecting each atom depict the nearest-neighbor component to the van der Waals interaction.  The rectangular dashed line shows a domain as a subset of atoms within the larger array.  
Right: Schematic of a usable domain abstracted as an ideal graph state.  Vertices are spin states, $\vert +\rangle$, and the lines depict entanglement with the CZ gate.  The arrows show the information flow in computational order parameters defining domain size.
}
\label{fig_domain}
\end{figure}

We demonstrate construction and benchmarking of large graph states on a Rydberg atom quantum analogue simulator, QuEra's Aquila \cite{Wurtz2023}, as depicted in Fig.~\ref{fig_domain}.  Aquila operates under the following analogue conditions \cite{Wurtz2023}: i) The location of up to 256 $^{87}\text{Rb}$ atoms are defined in the plane at the start of a run and held in place with optical tweezers for the entirety of the run,  ii) Pulsing ground-to-Rydberg transitions acts globally on all atoms at once, iii) Pairs of atoms in excited Rydberg states interact with a van der Waals interaction during the entire run,  and iv) Measurements of ground and Rydberg state populations are recorded globally for all atoms at the end of the run.  These analogue conditions have the advantage of parallel entanglement using an always on interaction but exclude many features of digital Rydberg-based devices, e.g., local control of one and two-atom gates \cite{Saffman2010,BROWAEYS2020a}.  Ref.~\onlinecite{Wurtz2023} recorded device parameters, local error benchmarks, and other details of the Aquila device.

We test our order parameters on Aquila. Active correction of weak errors would, in principle, imply a uniform finite-size scaling of computational order, and therefore computational domains of arbitrary size.  But as error probabilities increase above threshold we expect a transition to a regime with exponential decay of computational order scaling.  To see this consider one measurement per vertex where the probability of an uncorrelated error in a single measurement at a vertex is $\varepsilon_M$.  Non-local fidelities derived from $n$ measurements, and therefore computational order, will decrease as $\sim (1-\varepsilon_M)^n$ as we grow the number of measurements and vertices.   We parameterize the error probability in the Aquila device with similar scaling functions for computational order parameters.   Our scaling analysis indicates that errors limit domain sizes to be more than an order of magnitude smaller than those needed for thresholds implied by Eq.~\eqref{eq_error_threshold}.  Our domain-based benchmarks complement other recently implemented neutral atom benchmarking tools \cite{BLUVSTEIN2022a,Nikolov2023,SHAW2024,kornjaca2024}.

The paper is organized as follows. Section~\ref{sec_model_theory} discusses the model of atoms in the Aquila device and presents the method to create and characterize graph states.  In Sec.~\ref{sec_model} we discuss the Rydberg atom  quantum analogue simulator Hamiltonian.  Here we show how the always-on van der Waals interaction between atoms can be used to create graph states using parallel entanglement. We also quantify unitary errors inherent to the always-on protocol.  In Sec.~\ref{sec_string} we discus how to characterize graph states with stabilizer and string correlation functions.  In Sec.~\ref{sec_logical} we discuss measurement-based logic operations in the Clifford group as computational order parameters.  Section~\ref{sec_resutls_aquila} presents results from the Aquila device. In Sec.~\ref{sec_bell} we create two-particle entangled states on Aquila and test fidelity.  In Secs.~\ref{sec_aquila_stabailizer_strings} and ~\ref{sec_aquila_computational_order} we create graph states on Aquila and characterize them to estimate computational domain sizes.  Section~\ref{sec_aquila_error} discusses error sources.  We summarize in Sec.~\ref{sec_outlook}. 

\section{Model and Methods}
\label{sec_model_theory}

\subsection{Hamiltonian and Always-on Interaction} 
\label{sec_model}

We define the model and wavefunction of Rydberg atoms trapped in optical tweezer arrays.  Once $N$ atomic locations are fixed in the plane by optical tweezers, the time-dependent Hamiltonian is \cite{BROWAEYS2020a,Wurtz2023}: 
\begin{align}
\hat{H}(t)&=\hat{H}_0(t)+\hat{V} \nonumber \\
\hat{H}_0(t)&=\frac{\Omega(t)}{2}\sum_{i=1}^N\left[ e^{i\phi(t)} \vert g_i\rangle\langle r_i \vert + e^{-i\phi(t)} \vert r_i\rangle\langle g_i \vert \right] \nonumber\\
\hat{V}  & = \sum_{i<j}^N V(d_{ij}) \vert r_i \rangle \vert r_j \rangle \langle r_j \vert \langle r_i \vert, \nonumber
\end{align}
where $\vert g_i\rangle$ ($\vert r_i\rangle$) denotes the ground (excited Rydberg) state of the $i^{\text{th}}$ atom.  The location of each atom defines graph vertices in the real space plane.  $V(d_{ij})=C_6/d_{ij}^6$ are the inter-atom interaction strengths where, for the Aquila device, $C_6=5420503\hspace{0.05cm} \mu \text{m}^6 \text{rad}/\mu \text{s}$ and $d_{ij}$ is the planar inter-atom separation.  For each graph we use uniform distances, $d_{i,j}=d$, except at certain input vertices discussed in Sec.~\ref{sec_logical}.  The Rabi drive amplitude $\Omega(t)$ and the phase $\phi(t)$ can be pulsed during the run.  We set $\hbar=1$.  

\begin{figure}[t]
\begin{center}
\includegraphics[width=0.46\textwidth]{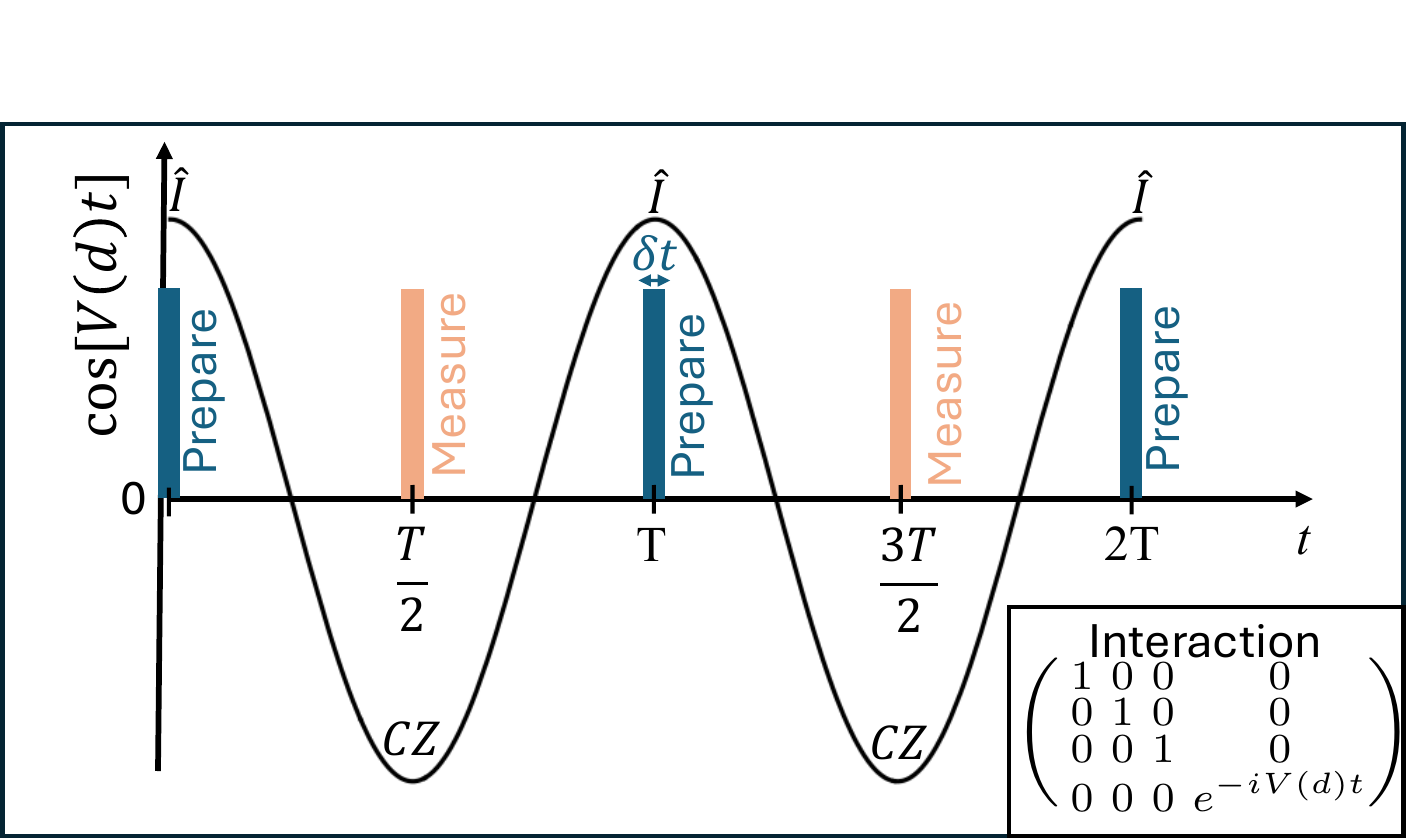}
\end{center}
\caption{Schematic of time scales in the always-on interaction protocol for creation of graph states with the controlled phase interaction between atoms. The solid line plots the real part of the phase in the interaction matrix as a function of time.  The slow interaction is a controlled phase operation between the atom pairs that approximates the two-spin identity gate, $\hat{I}$, at times that are even multiples of $T/2$ and the CZ gate at odd multiples, where  $T=2\pi/V(d)$.  The bars are schematics labeling times for short pulsing stages of duration $\delta t$.  The preparation stage uses a global Rabi-drive pulse to simultaneously rotate all atoms from $\vert g_i \rangle $ into the state $\vert +_i \rangle$.  The measurement stage similarly rotates all atoms and then measures $\{ \vert g_i \rangle, \vert r_i \rangle \}$ populations.  The inset shows the matrix for two-particle interaction time evolution,  $\exp{[-i \hat{V} t]}$, in the basis $\{ \vert g_1\rangle\vert  g_2\rangle,  \vert g_1 \rangle \vert r_2\rangle, \vert r_1 \rangle \vert  g_2\rangle, \vert r_1 \rangle \vert r_2\rangle\}$.  
}
\label{fig_always_on}
\end{figure}

The atomic states define a pseudospin. $\hat{H}$ can be written in terms of the usual Pauli matrices at atom $i$, $\hat{\bm{\sigma}}_i=(\hat{\sigma}^x_i,\hat{\sigma}^y_i,\hat{\sigma}^z_i)$, using $\hat{\sigma}^z_i=1-2\vert r_i\rangle\langle r_i\vert$ and $\hat{\sigma}^x_i=\vert g_i\rangle\langle r_i \vert + \vert r_i\rangle\langle g_i \vert $.  The eigenbasis of $\hat{\sigma}^x_i$ is then $\vert \pm_i\rangle \equiv (\vert g_i\rangle \pm  \vert r_i\rangle)/\sqrt{2}$.  Note that the van der Waals interaction between two atoms evolves as $\exp{[-i \hat{V} t]}$ to become a controlled phase operation:
\begin{align}
\widehat{\text{CP}}_{jk}(\theta) \equiv e^{-i\theta \vert r_j\rangle \vert r_k\rangle \langle r_k\vert\langle r_j \vert}, \nonumber 
\end{align}
where $\theta=tV(d_{jk})$ for atoms $j$ and $k$.  This operation becomes a 
CZ operation between pseudospins at times that are odd multiples of $\pi/V(d_{jk})$, e.g., when we have $\widehat{\text{CP}}_{jk}(\pi)$, as depicted in Fig.~\ref{fig_always_on}.

To see the role of an always-on interaction, we construct the wavefunction in the Schr{\"o}dinger picture.  
The exact many-body wavefunction evolves under the pulse profiles of $\Omega(t)$ and $\phi(t)$  as \cite{FETTER2012,Wurtz2023}:
\begin{align}
\vert \psi(t) \rangle&= \hat{U}_{\hat{H}_0}(t)   \hat{U}_{\hat{V}}(t) \prod_{i=1}^N \vert g_i\rangle
\label{eq_exact_wavefunction}
 \\
\hat{U}_{\hat{H}_0}(t)&=\mathcal{T} \exp{ \left[ -i\int_0^{t} 
 \hat{H}_0(\tau)  
 d\tau \right]} \\
\hat{U}_{\hat{V}}(t) &=\mathcal{T} \exp{ \left[ -i\int_0^{t} 
\hat{U}_{\hat{H}_0}^{\dagger}(\tau) \hat{V}(\tau)  \hat{U}_{\hat{H}_0}^{\vphantom{\dagger}}(\tau)
 d\tau \right]}, 
 \label{eq_state_propagator}
\end{align}
where $\mathcal{T} $ denotes time ordering.  For $\hat{V}=0$, properly designed pulse profiles $\Omega(t)$ and $\phi(t) $ can be used to simultaneously rotate all atoms in pseudospin space.  But for $\hat{V}\neq 0$, we must consider the simultaneous evolution of the single-atom control fields along with the interaction because $\hat{H}_0$ and $ \hat{V}$ do not commute in general.  We will see below that the Schr{\"o}dinger picture wavefunction conveniently separates the roles of $\hat{H}_0$ and $ \hat{V}$ whereas the separation is harder to see in the more conventional interaction picture wavefunction, $ \hat{U}_{\hat{V}}(t) \prod_{i=1}^N \vert g_i\rangle $. 

\begin{figure}[t]
\begin{center}
\includegraphics[width=0.46\textwidth]{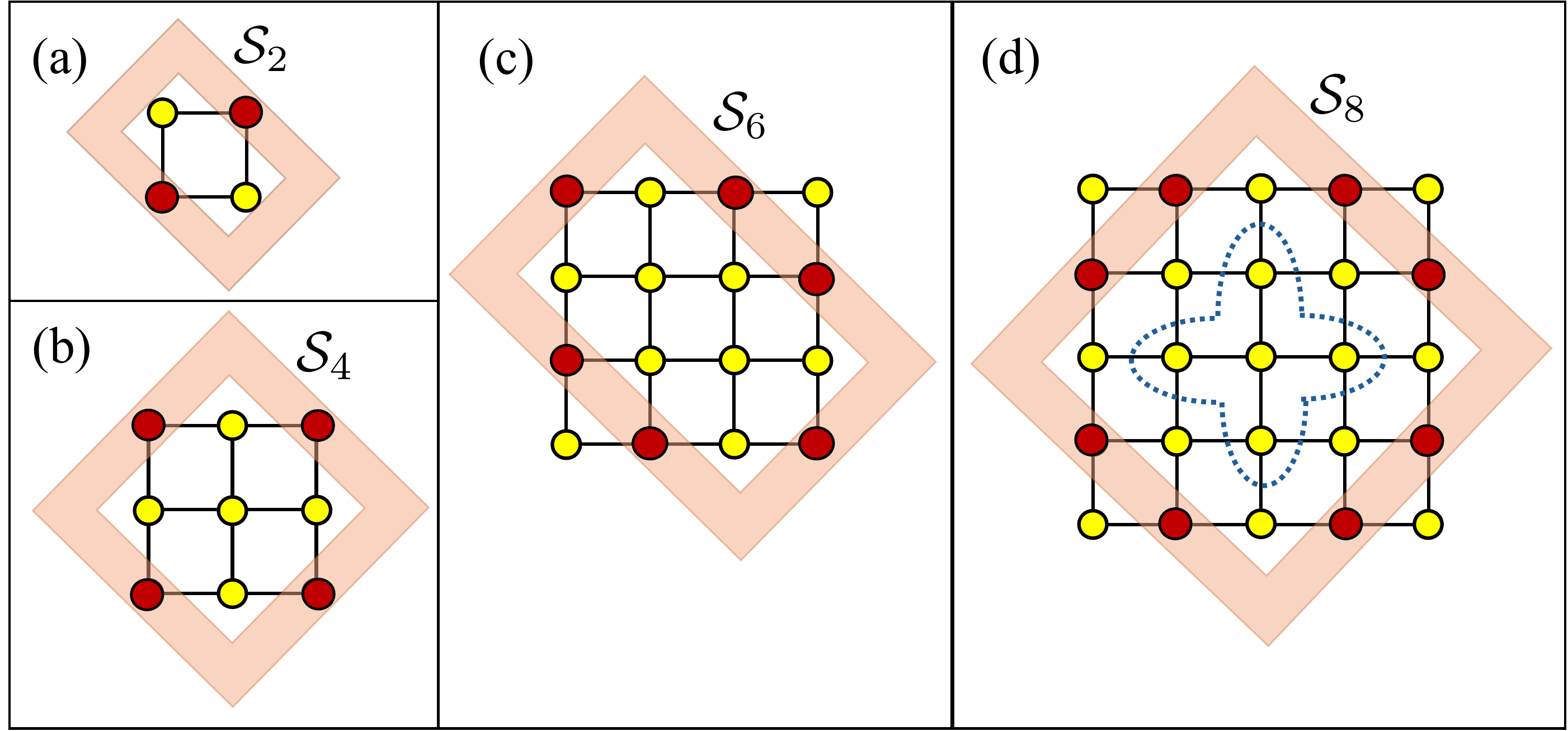}
\end{center}
\caption{ 
Schematics of example graph states created on Aquila.  Each circle represents an atom prepared in the $\vert + \rangle $ state.  Solid black lines depict entanglement with the always-on interaction approximation to the two-atom CZ operation. The shaded rectangles enclose atoms used to measure example strings, $\mathcal{S}_n$, defining non-local order parameters, Eq.~\eqref{eq_string_order_parameter}.  Atoms covered by $\mathcal{S}_n$ are shaded in red.   In (d), the dotted line about the center, $i_c$, encloses five atoms defining an example of one stabilizer operator that we measure to find $\langle \hat{S}_{i_c} \rangle$.
}
\label{fig_graphs_loops}
\end{figure}

We now discuss the always-on interaction as a route to create graph state edges with Rydberg atom arrays and estimate the impact of non-commutativity of $\hat{H}_0$ and $ \hat{V}$.  Fig.~\ref{fig_always_on} depicts subsequent rounds of preparation and measurement to create graph states and execute MBQC measurement protocols.  Ideally, preparation of $\prod_i \vert +_i\rangle$ allows the interaction to drive evolution into graph states for times at odd multiple of $T/2$. For infinitely narrow pulses and short-ranged interaction, this protocol leads to ideal graph states and can be efficiently scaled because it uses only global pulses.
For example, after the first measurement stage we expect a graph state:  
\begin{align}
\prod_{\langle j, k\rangle \in \mathcal{E}}\widehat{\text{CP}}_{j,k}(\pi) \prod_i \vert +_i\rangle,
\end{align}
where $\langle j, k\rangle$ indicates atom pairs defining graph edges $\mathcal{E}$.  Fig.~\ref{fig_graphs_loops}a-d depicts example vertex/edge combinations we build on Aquila.  

Perturbations arise due to finite pulse widths and long-range parts of the interaction.  These perturbations are known unitaries that can be incorporated into a definition of a weighted graph state \cite{Hein2006a}. Alternatively, known local unitary errors can be corrected with pulse engineering  \cite{Jones2003,Vandersypen2005} and perturbations due to the long-range interaction can be removed with additional pulsing \cite{Lee2016}.  Here we instead take these perturbations to be graph state preparation errors and quantify their impact.

We quantify the perturbation to the desired graph state caused by non-zero pulse widths evolving along with the interaction.  Consider the first preparation stage, $0\leq t \leq \delta t$, depicted schematically by the first vertical bar in Fig.~\ref{fig_always_on}.  Assume for simplicity that $\hat{H}_0(t)$ is a square pulse during the preparation stage.  To create a graph state we seek to first prepare  $\prod_i \vert +_i\rangle$ using 
just single-atom rotations: $\hat{U}_{\hat{H}_0}(\delta t) \prod_{i=1}^N \vert g_i\rangle$.  Ideally, interactions should entangle the state only after single-atom rotations are applied.  But the interaction perturbs the first preparation stage because it is always on.  To approximate the role of interactions, consider the first order Dyson series expansion for the propagator:
\begin{align}
\hat{U}_{\hat{V}}(\delta t)=
\hat{I}
-i\int_0^{\delta t} 
e^{i\hat{H}_0\tau} 
 \hat{V}
e^{-i\hat{H}_0\tau}
 d\tau +\mathcal{O}( [\delta t V(d) ]^2).
 \label{eq_approximate_propagator}
\end{align}
Substitution into Eq.~\eqref{eq_exact_wavefunction} shows that the second term in Eq.~\eqref{eq_approximate_propagator} induces a preparation stage correction (a unitary error) due to the always-on interaction, i.e., it causes deviations from $\hat{U}_{\hat{H}_0}(\delta t) \prod_{i=1}^N \vert g_i\rangle$.  Note that the correction to $\hat{I}$ scales as $\delta t V(d)$ in the wavefunction and therefore vanishes linearly for $\delta t\rightarrow 0$.  As mentioned above, pulse engineering \cite{Jones2003,Vandersypen2005,Qin2021} can, in principle, push the error to higher order in $\delta t$.  But as we will see quantitatively, the linear perturbation has a small effect on Aquila compared to other perturbations because we are able to choose single-atom pulse widths such that $\delta t /T$ is below 0.25 on Aquila.  A similar analysis applies to short pulses in the measurement stage for times near $T/2$.  We can therefore accurately characterize the weak unitary deviations of $\hat{U}_{\hat{V}}$ from $\hat{I}$ (CZ) during the preparation (measurement) stages.  In the remainder of this work we consider only one preparation and one measurement stage.

\subsection{Stabilizer and String Correlators in Graph States} 
\label{sec_string}

We now discuss the characterization of graph states as depicted in Fig.~\ref{fig_graphs_loops}a-d using expectation values of stabilizer and string correlation functions.  Measurement-based benchmarking can be implemented by testing the fidelity of local stabilizer operators and non-local string correlators defined by stabilizer products.  Products of stabilizer operators bound MBQC computational order because MBQC protocols are defined from combinations of stabilizer operator products. 

Graph states are eigenstates of stabilizer operators defined by \cite{Hein2006a}:
\begin{align}
\hat{S}_i=\hat{\sigma}_i^x\prod_{j\in \mathcal{N}}\hat{\sigma}^z_j,
\end{align}
where the product runs over the connections to vertices in the neighborhood of $i$, $j\in \mathcal{N}$, to define graph edges.   We may test the stabilizer average given by: $N_s^{-1}\sum_{i=1}^{N_s} \hat{S}_i$, where $N_s$ is the number of stabilizer operators fitting within the graph.  The dotted line in Fig.~\ref{fig_graphs_loops}d encircles five  vertices defining one example stabilizer operator.  Since graph states are unique eigenstates of $\hat{S}_i$ with eigenvalue $+1$, we expect to obtain unity on an ideal graph state.  $N_s^{-1}\sum_{i=1}^{N_s} \hat{S}_i$ therefore serves as an example graph state fidelity measure.

We can also use $\hat{S}_i$ to construct non-local string correlators that can be implemented with global pulses and measurements to reveal scaling for bounds of computational order.  Ideal graph states are also eigenstates of products of $\hat{S}_i$ defining non-local symmetries, e.g., $\prod_{j\in \mathcal{S}_n} \hat{\sigma}^x_j$, where $\mathcal{S}_n$ denotes certain strings of atoms containing $n$ measurements with one measurement per atom in the string.  Fig.~\ref{fig_graphs_loops} depicts example strings by rectangular loops passing though included atoms.  One can show that the eigenvalue of these example string operators is $+1$ for ideal graph states since all basis states contributing amplitude to the many-body wavefunction preserve parity of the string symmetry.

We define the expectation value of string symmetries as non-local order parameters obtained from $n$ measurements:
\begin{align}
\Theta_{n}\equiv \bigg \langle \hat{\mathcal{P}}^{+}_{\mathcal{S}_n} \bigg [ \prod_{j\in \mathcal{S}_n} \hat{\sigma}^x_j \bigg ] \hat{\mathcal{P}}^{+}_{\mathcal{S}_n} \bigg \rangle, 
 \label{eq_string_order_parameter}
 \end{align}
 where $\hat{\mathcal{P}}^{+}_{\mathcal{S}_n}$ are projectors onto basis states preserving the $+1$ string parity.  $\hat{\mathcal{P}}^{+}_{\mathcal{S}_n}$ defines a post selection protocol to set a normalization consistent with conventional fidelity measures: $\Theta_{n}=1$ for an ideal graph state and $\Theta_{n}=1/2$ for a product state randomized by errors.  We implement the projectors by selecting the measurement outcomes that preserve $+1$ parity.  This, without loss of generality, excludes the other half of all possible outcomes, i.e., $-1$ parity outcomes.  For example, in Fig.~\ref{fig_graphs_loops}a, the following string outcomes corresponding to amplitude in the two-spin wavefunction are kept: $\vert +\rangle \vert +\rangle$ and $\vert -\rangle \vert -\rangle$.  Outcomes corresponding to $\vert +\rangle \vert -\rangle$ and $\vert -\rangle \vert +\rangle$ are rejected.

   We can estimate scaling of $\Theta_{n}$ with $n$ using a classical single-atom error model.  Let $\varepsilon_L $ be the probability that an error causes a measurement of a single atom state $\vert \pm \rangle$ to flip to $\vert \mp\rangle$ anywhere along the length of $\mathcal{S}_n$.  Note that $ \Theta_{n}$ only changes with an odd number of flips in the eigenbasis of $\hat{\sigma}^x$.   The probability of flipping none or any even number of outcomes along the string follows from a binomial distribution:
\begin{align}
P_{\text{e}} (n)&=\sum_{k=0}^{k_{\text{m}}} \frac{n!}{(2k)!(n-2k)!} \varepsilon_L^{2k}(1-\varepsilon_L)^{n-2k} \nonumber \\
 &=\frac{1}{2}\bigg[ 1+(1-2\varepsilon_L)^{n} \bigg ],
               \label{eq_even_probability_error}
  \end{align}
where $k_{\text{m}}=n/2$ $[k_{\text{m}}=(n+1)/2]$ for $n$ even [odd].  $P_{\text{e}}(n)=1$ for $\varepsilon_L=0$.  For $\varepsilon_L>0$, $P_{\text{e}}(n)$ decays exponentially to $1/2$ as $n$ increases.  We take Eq.~\eqref{eq_even_probability_error} as an approximate error model for measurements of string correlators of graph states on Aquila that quantifies exponential scaling with a single fitting parameter, $\varepsilon_L$. 

\subsection{Computational Order Parameters} 
\label{sec_logical}

We now turn to non-local order parameters defined by measurement-based quantum logic operations in the Clifford group.  The order parameters rely on global measurements and, by construction, imply the efficacy to compute with the graph state.  Since these operations can be scaled in size (larger graphs produce the same logical operation), their fidelities can be used to estimate the scaling with $N$ and $n$.  

\begin{figure}[t]
\begin{center}
\includegraphics[width=0.46\textwidth]{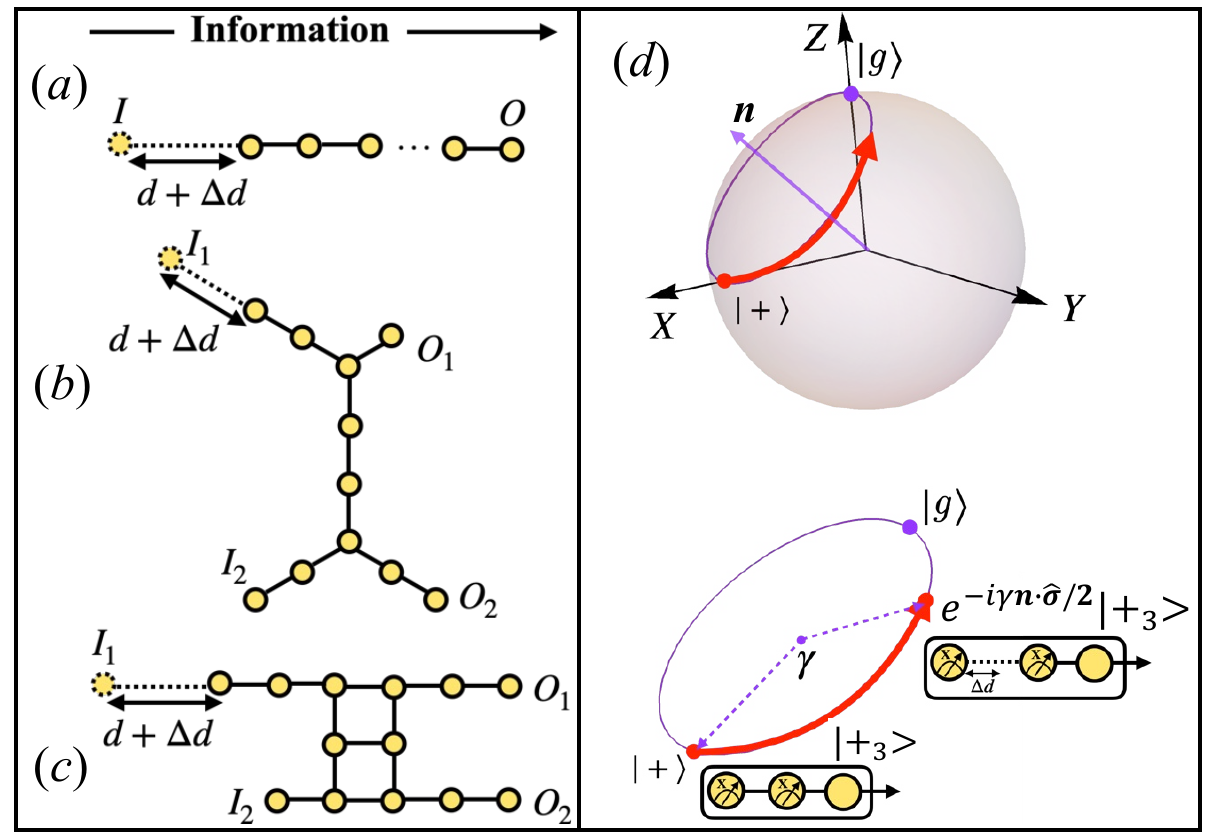}
\end{center}
\caption{ 
Measurements on the graphs in (a)-(c) define non-local computational order parameters based on MBQC quantum logic operations \cite{Raussendorf2003}. (a) Chain graph of atoms where global measurements along the pseudospin-$x$ eigenbasis teleport a quantum state from $I$ to $O$ to execute a logical identity operation.  (b) The measurements on this graph take input states at $I_1$ and $I_2$  and performs logical controlled-NOT to yield outputs at $O_1$ and $O_2$.  (c) The same as (b) but for a logical SWAP operation.  To prepare input states  in (a)-(c), the input atoms are translated by $\Delta d$.  (d) The red arrow depicts the impact of translation by $\Delta d$, measurements, and always-on interaction that takes the input $\vert +\rangle$ state at $I$ and transforms it so that the third atom is rotated by an angle $\gamma$ on the Bloch sphere about the unit vector $\bm{n}=(1,0,1)/\sqrt{2}$. The state at the third atom becomes the input for the rest of the graph.  
}
\label{fig_logical_bloch}
\end{figure}

We first review the measurement-based protocol to implement a logical identity operation \cite{Raussendorf2003,Tame2005,Qin2021,Qin2024b}.  Fig.~\ref{fig_logical_bloch}a depicts a chain graph that, upon global measurement, executes logical identity by teleporting the state encoded at input, I (at $i=1$), to output, O (at $i=N$), where vertices are labeled $i=1,2,...,N$, from left to right.  To see this, first consider an ideal chain graph state with vertices $i=2,3,...,N$: 
\begin{align}
\vert \psi_{\text{c}}\rangle=\prod_{j=2}^{N-1}\widehat{\text{CP}}_{j,j+1}(\pi) \prod_{i=2}^{N} \vert +_i\rangle,
\end{align}
where we excluded the left-most vertex at $i=1$.  We 
then encode information in the leftmost vertex: $\vert \psi_I\rangle$.  Entangling $\vert \psi_I\rangle$ with the rest of the graph leads to the state:
\begin{align}
\vert \psi_{\text{e}}\rangle = \widehat{\text{CP}}_{1,2}(\pi) \vert \psi_I\rangle \vert \psi_{\text{c}}\rangle,
\end{align}
which is now an $N$-vertex chain graph state with encoded information.  Measurements on $i=1,2,...,N-1$ move information in $\vert \psi_I\rangle $ from left to right along the graph.  One can show \cite{Raussendorf2003} that the output state encodes information at $i=N$ which is, up to a known unitary, the same as the input state:
\begin{align}
\vert \psi_{\text{O}}\rangle =  \hat{U}_{\Sigma} \hat{\mathcal{I}} \vert \psi_I\rangle, 
\label{eq_identity}
\end{align}
where $\hat{\mathcal{I}}$ denotes the logical measurement-based identity operation. $\hat{U}_{\Sigma}=(\hat{\sigma}^z)^{s_1+s_3+..}(\hat{\sigma}^x)^{s_2+s_4+..}$ is 
a byproduct unitary that feeds-forward measurement outcomes along the chain graph.  $s_i=0,1$ are the measurement outcomes from a projective $x$-measurement at vertex $i$.  

Equation~\eqref{eq_identity} defines an example measurement-based logical operation in the Clifford group.  It implies that ideal graph state chains allow measurements to teleport information along the chain. Eq.~\eqref{eq_identity} arises from the  underlying every-other vertex symmetry, $\prod_{i} \hat{\sigma}^x_{2i+1}$, in one-dimensional graphs and tests a necessary condition for computational power \cite{RAUSSENDORF2023}.  Also, the expected output in  Eq.~\eqref{eq_identity} does not depend on $N$ for an ideal graph state.  The fidelity to implement    
Eq.~\eqref{eq_identity} can therefore be used to benchmark imperfect graph states of any size.  Furthermore, only global measurements are needed.  Other measurement-based logic operations in the Clifford group can be used to benchmark graph states in a similar fashion. Figs.~\ref{fig_logical_bloch}b and ~\ref{fig_logical_bloch}c depict graphs that, upon global measurement, execute logical controlled-NOT and SWAP operations, respectively, by pushing information from inputs to outputs \cite{Raussendorf2003}.

It is not possible to directly tune the state of selected input atoms using local fields when operating under the constraints of the Aquila device discussed in Sec.~\ref{sec_introduciton}.  We now introduce a method to use graph geometry to effectively tune input states $\vert \psi_I\rangle$ without local fields.  We consider construction of a chain graph state of atomic pseudospins using the always-on interaction.  All atoms are first globally prepared in the $\vert +\rangle$ state.  Entangling atoms  $i=2$ to $N$ with the two-atom CZ operation approximates $\vert \psi_{\text{c}} \rangle$.  But to encode information at the left side of the chain graph state we entangle the $\vert +_1\rangle$ state with a controlled phase operation instead of CZ.  To achieve this with the always-on interaction between Rydberg atoms we translate the $i=1$ atom as depicted in Fig.~\ref{fig_logical_bloch}a.  The resulting encoded wavefunction is:
\begin{align}
\vert \tilde{\psi}_{\text{e}}\rangle = \widehat{\text{CP}}_{1,2}[TV(d+\Delta d)] \vert +_1\rangle \vert \psi_{\text{c}}\rangle.
\label{eq_psi_weighted}
\end{align}
Here, translation by $\Delta d$ encodes information that can be teleported from the left to the right side of the chain graph state.

To derive the output state from $\vert \tilde{\psi}_{\text{e}}\rangle$, consider just the first three atoms: $i=1,2,3$.   It is straightforward to show that after using measurement to trace out the first two atoms, the state at $i=3$ becomes our new input state for the rest of the graph:
\begin{align}
\vert \tilde{\psi}_I\rangle=e^{-i\gamma \bm{n} \cdot \hat{\bm{\sigma}}/2}\vert+_3\rangle,
\end{align} 
i.e., a state rotated about the unit vector $\bm{n}=(1,0,1)/\sqrt{2}$ by an angle $\gamma$ defined by:
\begin{align}
\tan\left(\frac{\gamma}{2}\right)=\sqrt{2}\tan\left[\frac{T}{2}\{V(d+\Delta d)-V(d)\}\right].
\label{eq_define_gamma}
\end{align} 
Figure~\ref{fig_logical_bloch}d depicts the impact of increasing $\Delta d$.  The remaining measurements along the chain graph state then teleport $\vert \tilde{\psi}_I\rangle$ from left to right along the graph to yield the output state:
\begin{align}
\vert \tilde{\psi}_O\rangle=\hat{U}'_{\Sigma}\hat{\mathcal{I}}\vert \tilde{\psi}_I\rangle,
\label{eq_rotated_output}
\end{align} 
where $\hat{U}'_{\Sigma}$ is a modified byproduct unitary (See Appendix~\ref{sec_app_byproduct}).  Eq.~\eqref{eq_rotated_output} shows that the information  encoded in the left side of the graph can be teleported to the right side of the graph.  

The above protocol combines graph geometry, the always-on interaction, and measurements to program input states.  Figs.~\ref{fig_logical_bloch}b and c depict intentional translation of one input atom for other graphs, where the first three atoms are similarly used to create input states for the rest of the graph.  These graphs depict translation of just one atom, but more sophisticated input geometries are possible.  In general, our method implies that weighted graph states, instead of local spin rotations, can be used to define input information for unweighted graph states. 

We can use the translation operation to build an order parameter with the chain graph in Fig.~\ref{fig_logical_bloch}a.  The $x$-measurements of the last (output) atom offers a non-local computational order parameter with $n=N$:
\begin{align}
Q_N(\gamma ) = \text{Tr}[\hat{\mathcal{M}}_x\hat{\rho}_O \hat{\mathcal{M}}_x^{\dagger}],
\label{eq_QN}
\end{align}
where $\hat{\mathcal{M}}_x$ denotes measurement along the pseudospin $x$-direction of the output atom and $\hat{\rho}_O$ is the density matrix of the output atom.  In the ideal case (i.e., in the absence of noise, long-range components to the interaction, and pulse widths) we expect projective $x$-measurements to move the rotated state, $\vert \tilde{\psi}_I\rangle$, to the output atom at O so that $Q_N(\gamma )$ becomes $q(\gamma)\equiv\vert \langle + \vert e^{-i\gamma \bm{n} \cdot \hat{\bm{\sigma}}/2}\vert +\rangle \vert^2=(3+\cos[\gamma])/4$.  Differences between the device results and the function $q(\gamma)$ serve as non-local graph state benchmarks. 

\section{Implementation on Aquila}
\label{sec_resutls_aquila}

\subsection{Two-Particle Entangled States } 
\label{sec_bell}

We first demonstrate the creation of two-atom superpositions to test the always-on approach.  The pulse sequence used to create the state is shown in Appendix~\ref{sec_app_pulse_profiles}. For $d=12.3\hspace{0.1cm}\mu$m the pulse sequence and always-on interaction approximates a Bell pair: $\vert B\rangle=(\vert g_1\rangle \vert g_2\rangle+\vert r_1\rangle \vert r_2\rangle)/\sqrt{2}$.  We find the overlap between $\vert B\rangle$ and the exact wavefunction, Eq.~\eqref{eq_exact_wavefunction}, to be 0.99, showing that here the correction term discussed in Sec.~\ref{sec_model} is small.  

\begin{figure}[t]
\begin{center}
\includegraphics[width=0.48\textwidth]{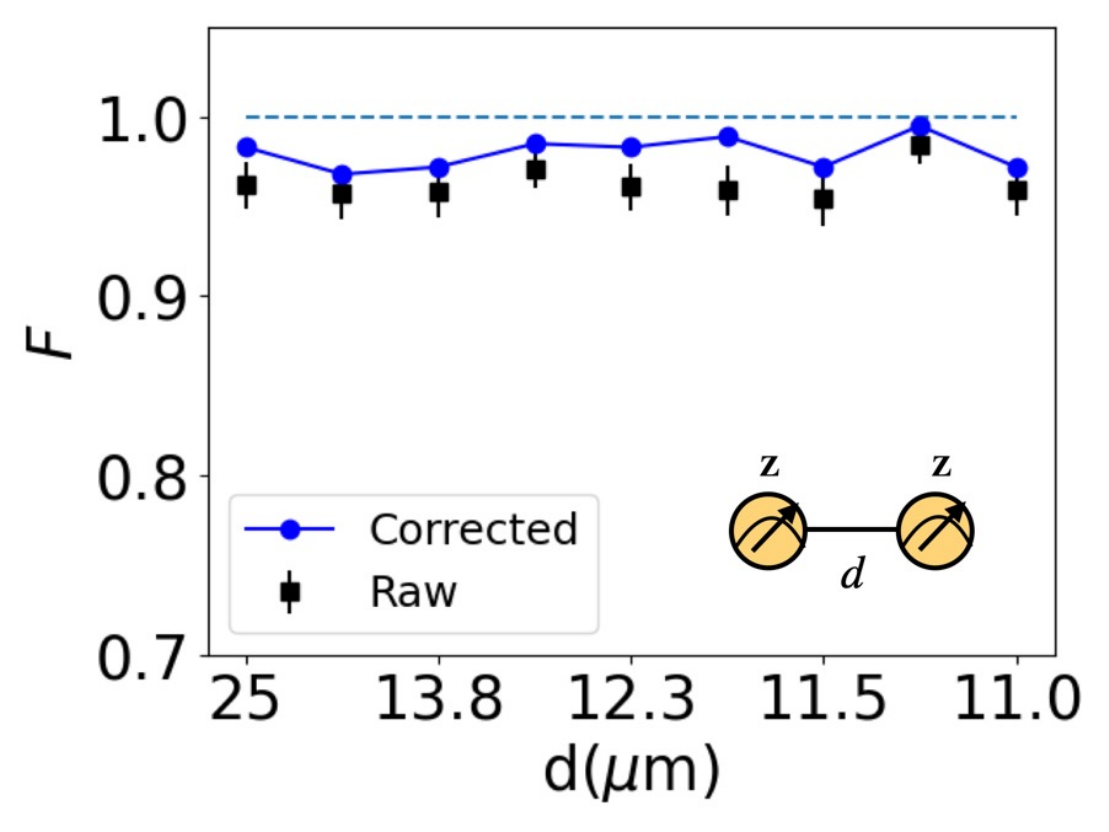}
\end{center}
\caption{ 
Fidelity of the two-atom state plotted as a function of inter-atom distance.  Squares plot the overlap of Aquila data with the exact wavefunction, Eq.~\eqref{eq_exact_wavefunction}, using
   100 shots. The circles plot the same but for corrected data that account for known \cite{Wurtz2023} measurement bias (See Appendix~\ref{sec_app_mitigation}).  The Bell state is expected at $d=12.3\hspace{0.1cm}\mu$m.  The solid line is a guide to the eye. }
\label{fig_pair_fidelity}
\end{figure}

 We also test the implementation of other two-particle entangled states 
 on Aquila.  The squares in Fig.~\ref{fig_pair_fidelity} plot a fidelity defined as the overlap of the exact wavefunction, Eq.~\eqref{eq_exact_wavefunction}, and the Aquila device.  The overlaps  demonstrate the high accuracy of the always-on protocol.  The error bars are computed using the delete-1 jackknife resampling method to obtain the standard error.  We choose enough shots to ensure convergence  (Appendix~\ref{sec_app_shot_scaling}). 
 
 Measurements on the Aquila device have an $\approx 8\%$ measurement bias of the $\vert r \rangle $ state \cite{Wurtz2023} which we have checked with 1000 single-atom shots. 
 This is the largest source of error.  We correct for measurement bias using the stochastic $\mathbb{T}$-matrix method \cite{MACIEJEWSKI2020,GELLER2020,CAI2023a} described in Appendix~\ref{sec_app_mitigation}  (Remaining sources of error will be discussed in Sec.~\ref{sec_aquila_error}).  The circles in Fig.~\ref{fig_pair_fidelity} plot the fidelity using measurement-corrected data.   We find that that the two-particle entangled states are created to within an overlap of $\approx 0.975$ after accounting for finite pulse widths and measurement bias.

\subsection{Estimates of Stabilizer and String Correlators on Aquila} 
\label{sec_aquila_stabailizer_strings}

We use Aquila to construct the graph states shown in Fig.~\ref{fig_graphs_loops} and benchmark them using the stabilizer and string correlators.   We use the pulse sequence discussed in Appendix~\ref{sec_app_pulse_profiles}. We first measure the stabilizer sum, $N_s^{-1}\sum_{i=1}^{N_s} \hat{S}_i$, in the pseudospin-$z$ eigenbasis using 1000 shots.  For $N=36$, we find the stabilizer average to be 0.985(6) which is consistent with preparation of a graph state.  An average near unity is a necessary but not sufficient demonstration of computational order with graph states.  The measurements used here are global and cannot be used with $\hat{S}_i$ to distinguish between graph states and product states (a combination of local $x$ and $z$ measurements are needed \cite{Jandura2024}).  Also, $\hat{S}_i$ is a local (5-atom) operator that does not scale with $N$ or $n$. 

\begin{figure}[t]
\begin{center}
\includegraphics[width=0.46\textwidth]{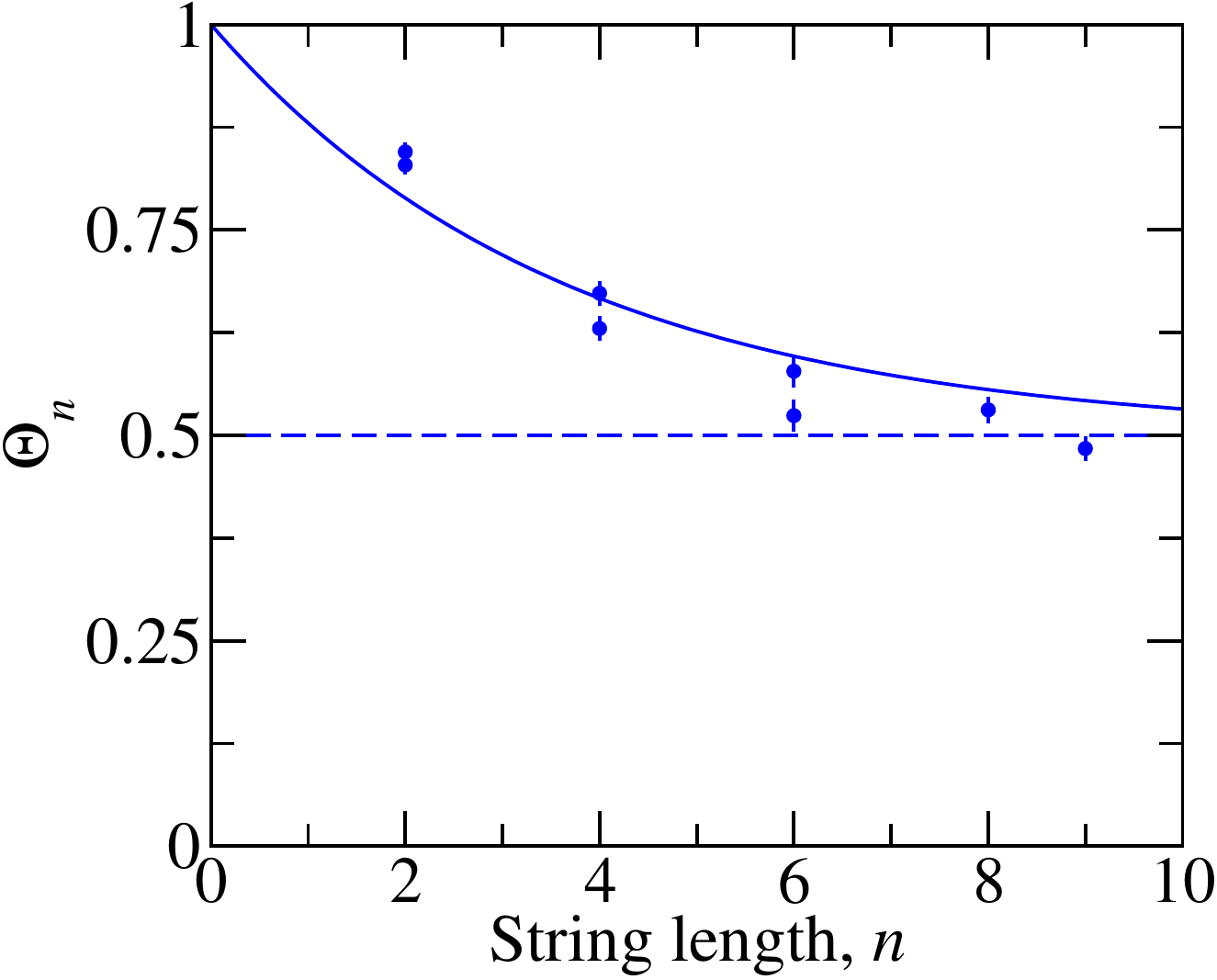}
\end{center}
\caption{ The circles plot estimates of string order, Eq.~\eqref{eq_string_order_parameter}, as a function of string length taken with 1000 shots on Aquila after creating graph states.  The data are corrected for measurement bias (Appendix~\ref{sec_app_mitigation}).  Graphs with up to 36 atoms were used and example strings are plotted in Fig.~\ref{fig_graphs_loops}.  The solid line shows a single parameter fit using Eq.~\eqref{eq_even_probability_error} to find $\varepsilon_L=0.12(1)$.  The horizontal dashed line plots the threshold for a randomized product state. 
}
\label{fig_loop_scaling}
\end{figure}

 We also used Aquila to infer $\Theta_{n}$ for various strings.  The circles in Fig.~\ref{fig_loop_scaling} show how the order parameter decreases with increasing string length.  The solid line in Fig.~\ref{fig_loop_scaling} shows a standard fit with Eq.~\eqref{eq_even_probability_error} yielding $\varepsilon_L=0.12(1)$.  We have therefore parameterized the exponential decrease in string correlations in these graph states.  The threshold $\Theta_n>1/2$ establishes that fewer than $\approx 9$ measurements on 9 atoms can be used to distinguish a graph state from a randomized product state.  These results therefore provide an upper bound on the domain size across which MBQC can be performed due to errors.

\subsection{Estimates of Computational Order Parameters on Aquila} 
\label{sec_aquila_computational_order}

\begin{figure}[t]
\begin{center}
\includegraphics[width=0.4\textwidth]{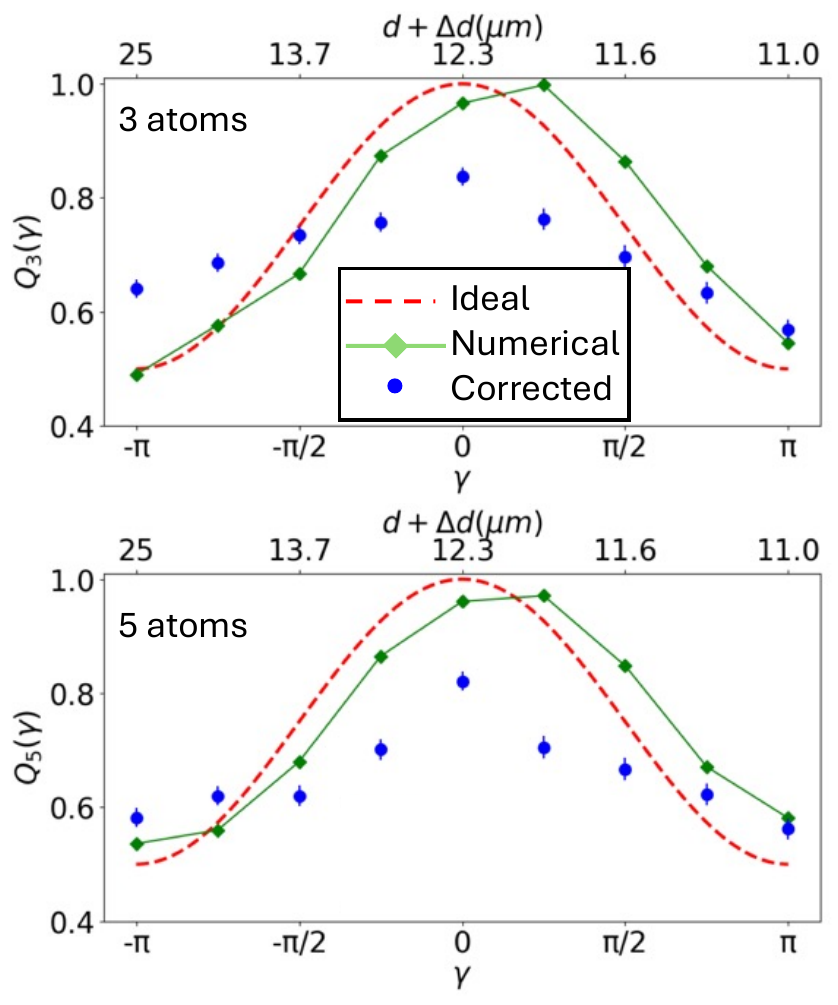}
\end{center}
\caption{ 
Example results for teleportation along chain graphs with 3 and 5 atoms on the top and bottom, respectively.  The $x$-axis is the rotation angle implemented by translating the first atom in the chain graph by $\Delta d$ (See Figs.~\ref{fig_logical_bloch}a and ~\ref{fig_logical_bloch}d).  The red dashed line shows the ideal case,  $q(\gamma)=(3+\cos[\gamma])/4$, expected for infinitely narrow pulses, only nearest-neighbor interaction, and no other sources of error.  The solid green line shows results from a numerical simulation that includes errors from finite pulse widths.  The circles are results from the Aquila device with 1000 shots, corrected for measurement bias (Appendix~\ref{sec_app_mitigation}). }
\label{fig_cluster_state_fidelity}
\end{figure}

We now use our protocol to estimate computational order parameters of states engineered on Aquila.  We build graph states with various shapes and sizes.  To test computational power we benchmark the states with the computational order parameters discussed in Sec.~\ref{sec_logical}. 

We first create the chain graph state, Fig.~\ref{fig_logical_bloch}a.  While two qubit fidelities are high, we must benchmark the computational power of the many-body wavefunction as system size is scaled.  We use $Q_N(\gamma )$ [See Eq.~\eqref{eq_QN}] and compare with the ideal case, $q(\gamma)$. 

Figure~\ref{fig_cluster_state_fidelity} shows measurement results along with results from numerical simulation and $q(\gamma)$.  The horizontal shift between $q(\gamma)$ and the numerical simulation result shows the impact of finite pulse widths.  The overall trend in the measurements shows evidence for teleportation of the rotated input state across the entire chain graph.   

We now turn to scaling estimates of the teleportation fidelity $Q_N(0)$ with $N$.  We expect $Q_N(0)=1$ for no error and $Q_N(0)=1/2$ for large error.  As in Sec.~\ref{sec_string}, we consider an uncorrelated measurement error model to construct a scaling function.  We assume that measurement errors change values of $s_i$ in the byproduct operator.  Applying a byproduct operator with incorrect $s_i$ to $\vert \pm_i \rangle$ will only cause an error if $s_i$ is changed an odd number of times on atoms $i=1,3,5,...,N$, for $N$ odd.  We therefore see that an odd number of measurement errors must occur on the $n_O\equiv(N+1)/2$ atoms (at every other atom) to cause an error, consistent with the every-other vertex symmetry, $\prod_{i} \hat{\sigma}^x_{2i+1}$.  This motivates a scaling fidelity $P_e(n_O)$ that parametrizes the total probability for only an even number of measurement errors to occur on $n_O$ atoms.  Appendix~\ref{sec_app_trajectory} discusses an alternative scaling function based on amplitude damping in a trajectory model \cite{CARLO2004}. 

Teleportation fidelity, and therefore $Q_N(0)$, has a threshold.  A teleportation fidelity above $2/3$ guarantees teleportation across a quantum channel \cite{Horodecki1999,Braunstein2001,Bose2003,Qin2021}.  A fidelity below $2/3$ leaves no guarantee that the graph contains a quantum channel.  We use this threshold to define domain size.

\begin{figure}[t]
\begin{center}
\includegraphics[width=0.43\textwidth]{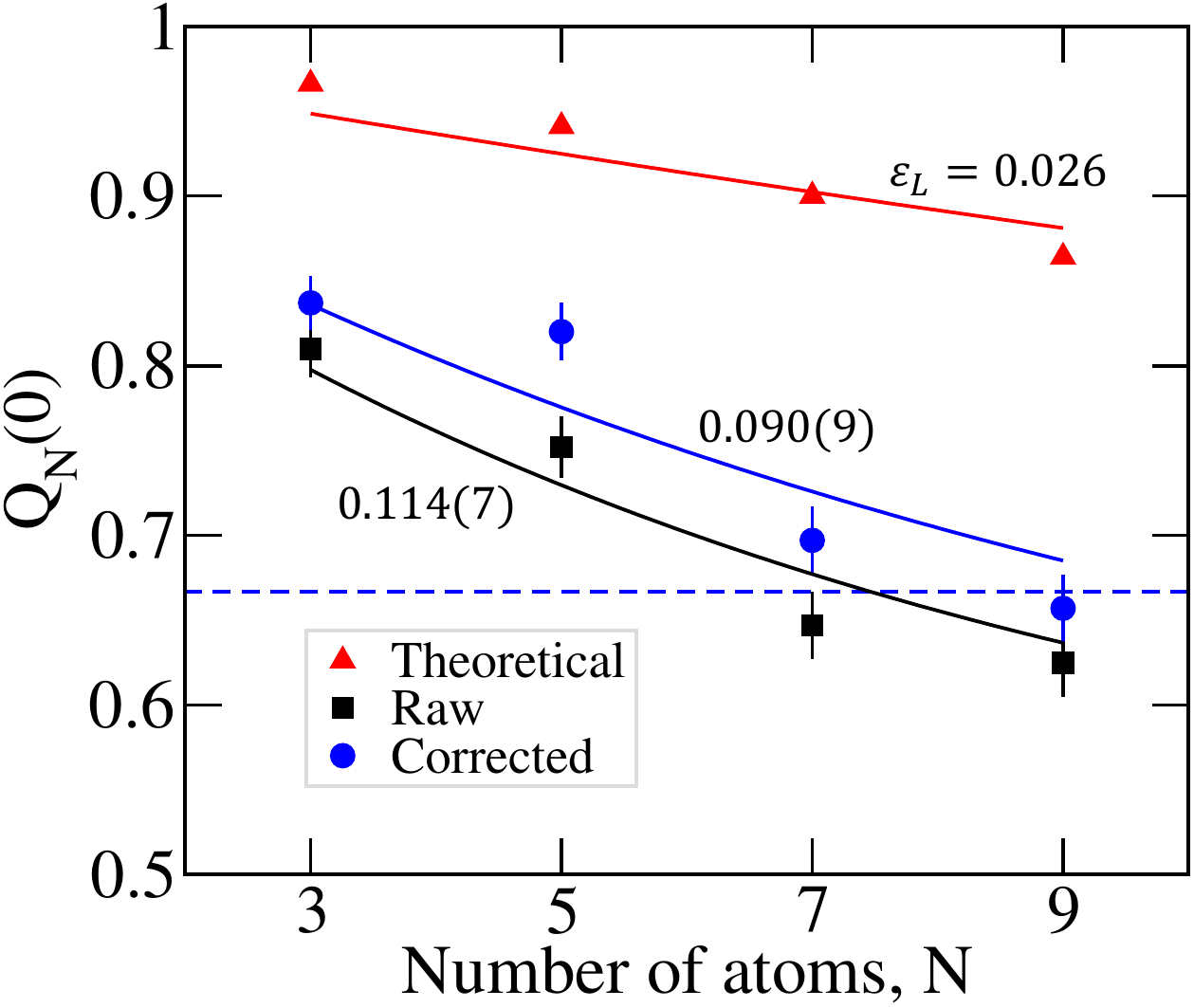}
\end{center}
\caption{ 
The teleportation fidelity along the chain graph as a function of the number of atoms.  The triangular data points are a theoretical prediction using the exact wavefunction, Eq.~\eqref{eq_exact_wavefunction}, that includes errors due to the finite pulse widths (See Appendix~\ref{sec_app_pulse_profiles} for the pulse shapes) and long-range parts of the interaction.  The squares are raw data from Aquila and the circles plot data corrected for measurement bias. The solid lines are single parameter fits with $P_e(n_O)$ to obtain $\varepsilon_L$.  The horizontal dashed line depicts the 2/3 threshold below which we cannot guarantee teleportation along a quantum channel.  
}
\label{fig_scaling}
\end{figure}

Figure~\ref{fig_scaling} shows the scaling in measurement-based teleportation we obtain from the Aquila device.  The triangles show a theoretical expectation for $Q_N(0)$ that incorporate the errors due to the finite pulse width and the long-range part of the interaction using the exact wavefunction.  The squares show the raw data from Aquila and the circles show the data corrected for measurement bias.  The solid lines show single parameter fits with $P_e(n_O)$.  A standard fit to the corrected data yields $\varepsilon_L=0.090(9)$, consistent with the string error probability obtained in Sec.~\ref{sec_string} and the error model in Appendix~\ref{sec_app_trajectory}.  The corrected data fall below 2/3 for chain graphs with $N\gtrsim 9$.  This is also consistent with results from Sec.~\ref{sec_string} showing that string lengths $n\gtrsim 9$ had a saturated string order parameter.  We therefore conclude that our implementation of graph states on Aquila support computational domains smaller than $\approx 9$ atoms and measurements.  These domain sizes imply that we can also demonstrate the MBQC controlled-NOT operation.  We were able to demonstrate the MBQC controlled-NOT operation (See Fig.~\ref{fig_logical_bloch}b) to within an accuracy of 68(2)\% using 1000 shots after correcting for measurement bias.  

Figure~\ref{fig_logical_bloch}c depicts the minimal graph needed to implement the measurement-based logical SWAP operation where input states are pushed by measurement, from left to right, to eventually swap at the output.  The six central vertices in the SWAP operation can be scaled up to rectangular arrays of arbitrary size.  This makes the MBQC SWAP operation the perfect candidate order parameter to test two-dimensional domain size scaling.  The graph shown in Fig.~\ref{fig_logical_bloch}c is too large to implement given the current error budget on Aquila.

\subsection{Error Budget on Aquila} 
\label{sec_aquila_error}

Several different error sources on Aquila impact the computational order parameters. The scaling estimate of $\varepsilon_L$ in Fig.~\ref{fig_scaling} can be thought of as an aggregate parameterization of many error sources.  
We interpret the accumulation of error using the known hierarchy of errors on Aquila \cite{Wurtz2023}.  The largest single source of error is the measurement bias, which we mostly account for by using measurement-corrected Aquila data to find $\varepsilon_L=0.090(9)$.  Error also arises from (unitary) perturbations from the finite pulse width and the long-range part of the interactions.  These errors are included in the estimate of error in $\varepsilon_L=0.026$ in Fig.~\ref{fig_scaling}.  We conclude that the difference between the corrected and theoretical fits, $\Delta \varepsilon_L = 0.064(9)$, can be understood as due to remaining sources of error.  

  Fluctuations in positions of the atoms due to thermal motion and tweezer location errors: $\delta d \approx 0.25 \hspace{0.1cm}\mu$m \cite{Wurtz2023}, are another possible source of error.  Modeling the fluctuations as fast harmonic oscillations in each planar direction, the interaction during the pulse experiences an averaged perturbation of $\approx 36 C_6 \delta d /d^7$ which we have checked changes $\varepsilon_L$ by less than 0.01.  We conclude that atom location errors are not a dominant contributor to $\Delta \varepsilon_L$.

 The largest contributions to $\Delta \varepsilon_L$ 
 are most likely due to an accumulation of single-atom decoherence effects.  Single-atom decoherence has been characterized by $T_2^*$ Ramsey measurements where it was found \cite{Wurtz2023} that $\vert g \rangle $-to-$\vert g \rangle $ transitions driven by conventional Ramsey pulses have an error probability of $\approx 0.07$ after $t=2 \mu$s.  (Possible sources include detuning from laser and Doppler effects, as well as the hybridization and decay with the intermediate state \cite{Wurtz2023}.) This error probability is consistent with our estimate of $\Delta \varepsilon_L = 0.064(9)$.  We therefore conclude that single-atom decoherence effects are the remaining dominant sources of error in our scaling estimate of $\varepsilon_L$.  
 
\section{Discussion and Outlook}
\label{sec_outlook} 

Non-local order parameters based on stabilizers and MBQC logic operations in the Clifford group can characterize graph states of any size.  These order parameters test the viability of graph state entanglement and measurements to execute measurement-based logic operations.  Finite-size scaling of the order parameters define the error-limited domain size over which graph states can be used for computation provided we define order parameter thresholds.  We imposed a threshold to string order to distinguish it from a classical string, $\Theta_n>1/2$, and a bound for guaranteed teleportation along a quantum channel, $Q_N(0)>2/3$.  

We demonstrated graph state construction on QuEra’s Aquila device.  An always-on interaction enables parallel entanglement.  Two types of unitary perturbations cause deviations from ideal graph states.  First, the non-commutativity of Rabi-drive pulses and the van der Waals interaction leads to unitary perturbations quantified with a Dyson series expansion.  Second, long-range components to the interaction also lead to perturbations.  We interpreted these quantifiable perturbations as state preparation errors.  They are relatively small and can be corrected with pulse engineering using modified control fields, e.g., refocusing schemes \cite{Jones2003,Vandersypen2005,Qin2021} and Floquet methods \cite{Lee2016}.  Alternatively, these intrinsic perturbations can be kept as useful \cite{Anders2006} features in defining weighted graph states \cite{Hein2006a}.

Scaling of computational order parameters tested the accumulation of errors.  Unitary errors arose from the always-on approach.  Other error sources included measurement errors and single-atom decoherence effects.  The ideal operating regime for the always-on approach is one where $T$ is much longer than pulse durations but much shorter than decoherence times.  We found that errors limit computational order to  domains containing fewer than $\approx 9$ measurements on as many atoms on Aquila using both the string order and the teleportation thresholds to define domain size.  Errors therefore limit the graph state domain size for MBQC to be less than 9 vertices.  

We can compare the domain sizes found here with an ideal case.  Assuming only measurement errors such that $\epsilon_{\text{M}}\ll 1$,  we set  $P_e(n_O)$ to $2/3$ to find domain sizes of $\sim \ln[3]/\epsilon_{\text{M}}$ vertices.  The error probability threshold given by Eq.~\ref{eq_error_threshold} thus leads to domains of $145$ vertices.  Computational domains of at least $145$ vertices must therefore be embedded inside larger graph states to correct measurement errors.  We therefore conclude that domains size found on Aquila are more than an order of magnitude smaller than the minimum needed to implement active error correction on a graph state. 

Quantum graph states are ideally constructed from parallel application of entangling gates.  We showed that Aquila allows parallel entangling gates using the always-on Rydberg interaction.  But this analogue mode of operating currently has limited coherence.  Significant improvements in scaling of computational order can be obtained by implementing high accuracy features of neutral atom digital quantum computing, e.g., rapid/high fidelity measurements implemented in optical lattices \cite{Wang2015} and those demonstrated in recent Rydberg atom tweezer arrays  \cite{Saffman2010,LABUHN2016,BROWAEYS2020a,EBADI2021a,Scholl2021,CONG2022,Deist2022,Graham2022,BLUVSTEIN2022a,Lis2023,Nikolov2023,Ma2023,Singh2023,Norcia2023,EVERED2023a,SHAW2024,BLUVSTEIN2024,MANETSCH2024,Flavien2024,Norcia2024}.

\begin{acknowledgments}
We thank Hrushikesh Sable for insightful discussions.  We acknowledge support from AFOSR (FA2386-21-1-4081, FA9550-23-1-0034, FA9550-19-1-0272) and ARO W911NF2210247.  
\end{acknowledgments}

\appendix

\section{Pulse Profiles}
\label{sec_app_pulse_profiles}

\begin{figure}[t]
\begin{center}
\includegraphics[width=0.41\textwidth]
{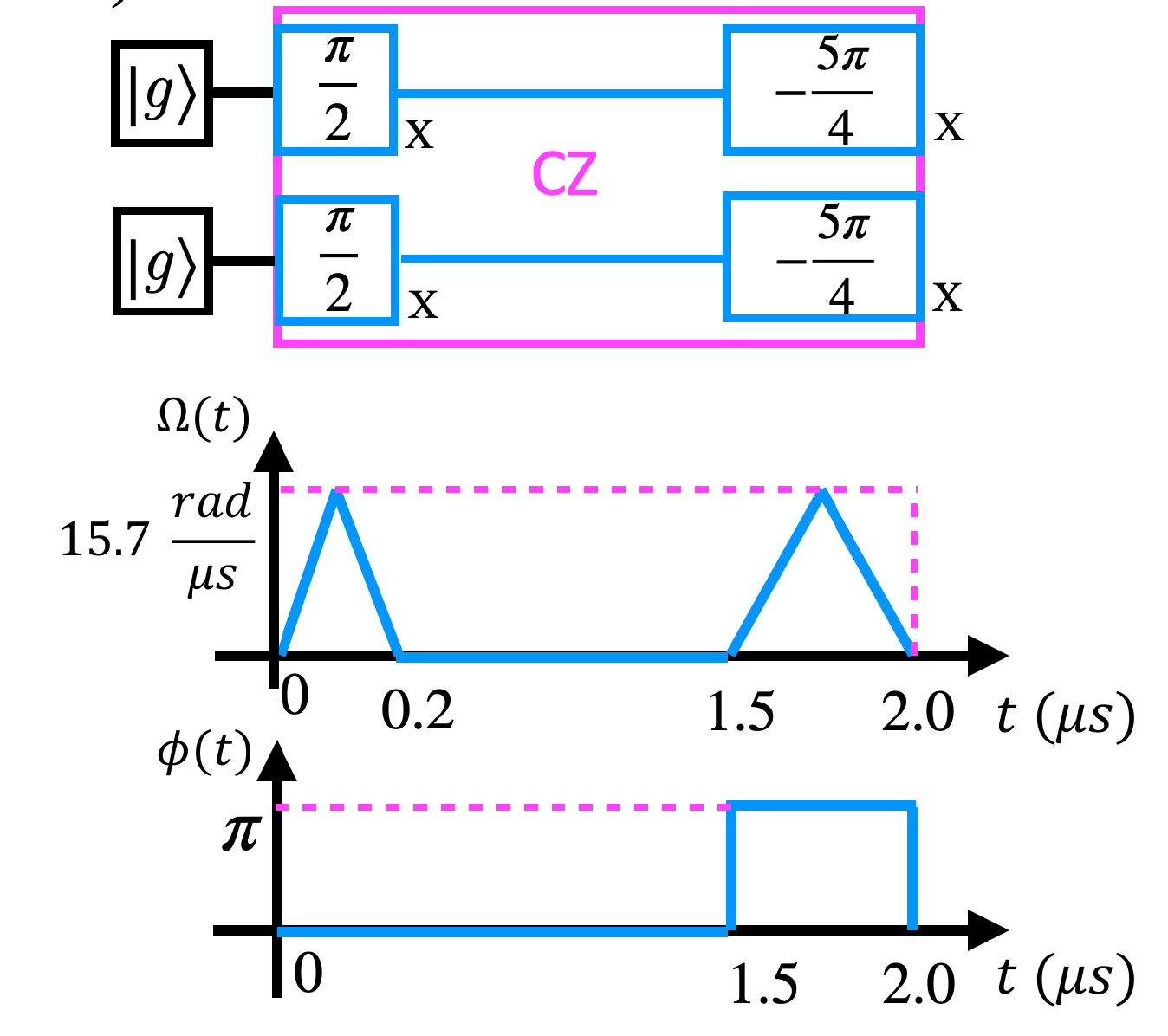}
\end{center}
\caption{ 
Pulse sequence used to create two-particle entangled states.  Atoms begin in $\vert g \rangle$.   
   In the absence of interaction or other perturbations, the ideal preparation (measurement) stage rotates both atoms around the pseudospin $x$-axis by $\pi/2$ ($-5\pi/4$). The always-on interaction perturbs the otherwise ideal pulses. The interaction eventually approximates the CZ operation after $2\hspace{0.1cm}\mu$s.  Measurements are performed at the end. For these parameters, the final state approximates a Bell pair state, $\vert B\rangle$, when atoms are separated by $d=12.3\hspace{0.1cm}\mu$m.  }
\label{fig_pair_circuit}
\end{figure}

This section presents the pulse profiles used to create quantum states on the Aquila device.  Figure~\ref{fig_pair_circuit} shows the circuit used to construct two-atom Bell pairs with the always-on interaction.  The preparation stage is in the first 0.2$\mu$s and the measurement stage is in the last 0.5$\mu$s.  Several pulse shapes were tested. Triangular pulse shapes for $\Omega(t)$ yielded the highest fidelity under the constraint of narrow pulse widths. 

 \begin{figure}[t]
\begin{center}
\includegraphics[width=0.4\textwidth]{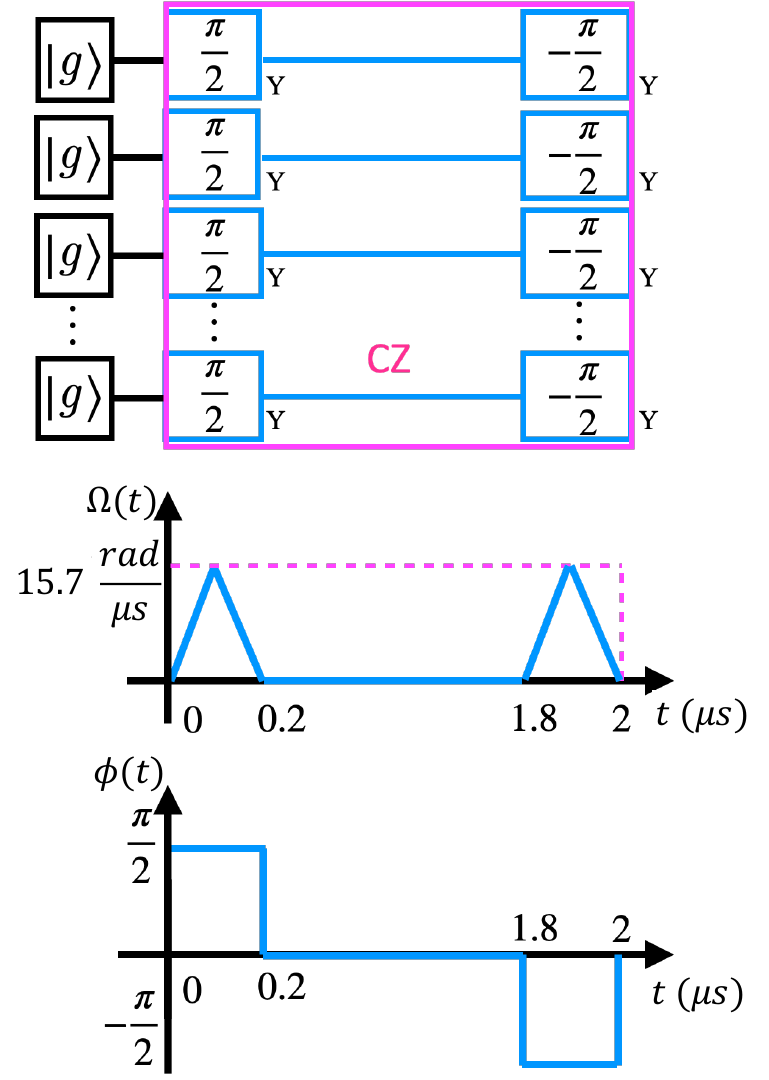}
\end{center}
\caption{ 
The pulse sequence used to create graph states where the preparation (measurement) stage rotates all atoms around the pseudospin-$y$ axis by $\pi/2$ ($-\pi/2$).  The always-on interaction approximates the CZ operation between nearest neighbors with $d=12.3 \hspace{0.1cm} \mu$m in the graph after $2\hspace{0.1cm}\mu$s.  Measurements are performed at the end.  }
\label{fig_cluster_state_circuits}
\end{figure}

Figure~\ref{fig_cluster_state_circuits} shows the pulse sequence to create chain graph states.  Here the always-on interaction simultaneously entangles all atoms.  The preparation stage is in the first 0.2$\mu$s and the measurement stage is in the last 0.2$\mu$s.  All other graph states are created with this pulse sequence as well.

\section{Error bars and shot number}
\label{sec_app_shot_scaling} 

This section describes how the shot number on Aquila was selected.  The number of shots required to achieve sufficiently small error bars increases with $n$.  The two-atom data presented in Fig.~\ref{fig_pair_fidelity} required only 100 shots to achieve convergence within tolerances set by small error bars.  But for larger graphs we increased the shot number.

\begin{figure}[t]
\begin{center}
\includegraphics[width=0.43\textwidth]{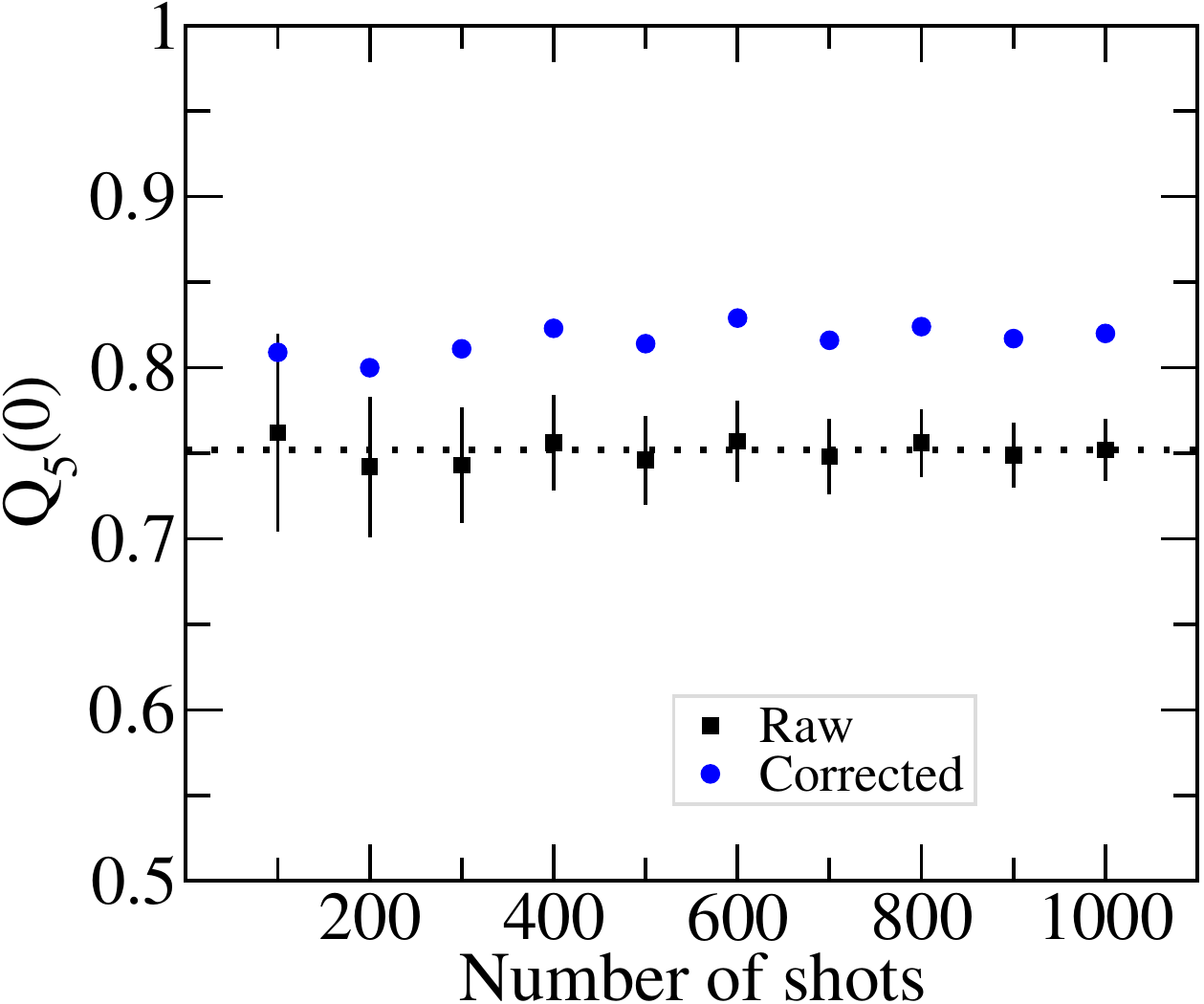}
\end{center}
\caption{An example convergence check plotting the estimate of $Q_5(0)$ [See Eq.~\eqref{eq_QN}] obtained from Aquila (squares) as a function of the number of shots used in computing averages.  The horizontal dotted line shows the average of raw data obtained with 1000 shots, 0.752, used in the main text. The circles plot the measurement corrected averages.  The error bars for the corrected data are very close to the error bars shown for the raw data and have been omitted for clarity. 
 }
\label{fig_app_shot_scaling}
\end{figure}

Figure~\ref{fig_app_shot_scaling} shows an example convergence check for data taken on Aquila with $N=5$ and $n=5$ using the pulse profile from Appendix~\ref{sec_app_pulse_profiles}.  Here we see no statistically significant deviation for large numbers of shots and that 1000 shots are more than sufficient for convergence to within small error bars.  The main text presents averages using enough shots to ensure convergence.

\section{Mitigation of Measurement Bias}
\label{sec_app_mitigation}

\begin{figure}[t]
\begin{center}
\includegraphics[width=0.46\textwidth]{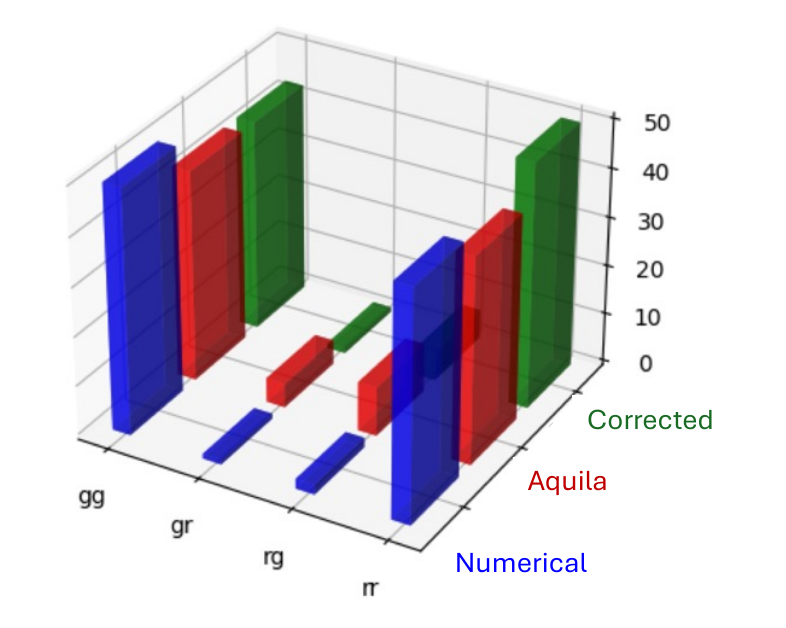}
\end{center}
\caption{ 
Example data for two atoms prepared to create a Bell state using the always-on interaction with the pulse protocol discussed in Appendix~\ref{sec_app_pulse_profiles} with 100 shots and $d=12.3\mu$m.  The histograms plot the number of outcomes.   The corrected data plot $\bm{m}_{\text{corr}}$ and show an improvement over the raw data from Aquila.  The numerical simulation includes corrections due to the finite pulse width during the always-on interaction. }
\label{fig_sup_histogram_bell}
\end{figure}

This section describes an error mitigation protocol for uncorrelated classical measurement error.  In this section we assume $n=N$. On the Aquila processor, the largest error source is a measurement bias where a fraction $\epsilon_M\approx0.08$ of measurements of the state $\vert g \rangle$ are incorrect and should be $\vert r \rangle$ \cite{Wurtz2023}.  We correct these measurement errors using the $\mathbb{T}$-matrix method \cite{MACIEJEWSKI2020,GELLER2020,CAI2023a}.  The left-stochastic error matrix $\mathbb{T}$ maps ideal measurements of $\vert g\rangle$ and $\vert r \rangle$ to noisy output via: $\bm{m}_{\text{meas}}=\mathbb{T}\bm{m}_{\text{corr}}$, where $\bm{m}_{\text{corr}}$ and $\bm{m}_{\text{meas}}$ are vectors containing normalized measurement counts for each outcome for the corrected and measured set, respectively.  By inverting this matrix, we correct measured data to account for measurement bias using: $\bm{m}_{\text{corr}}=\mathbb{T}^{-1}\bm{m}_{\text{meas}}$.  Data labelled ``corrected" plot results using $\bm{m}_{\text{corr}}$.  

\begin{figure}[t]
\begin{center}
\includegraphics[width=0.46\textwidth]{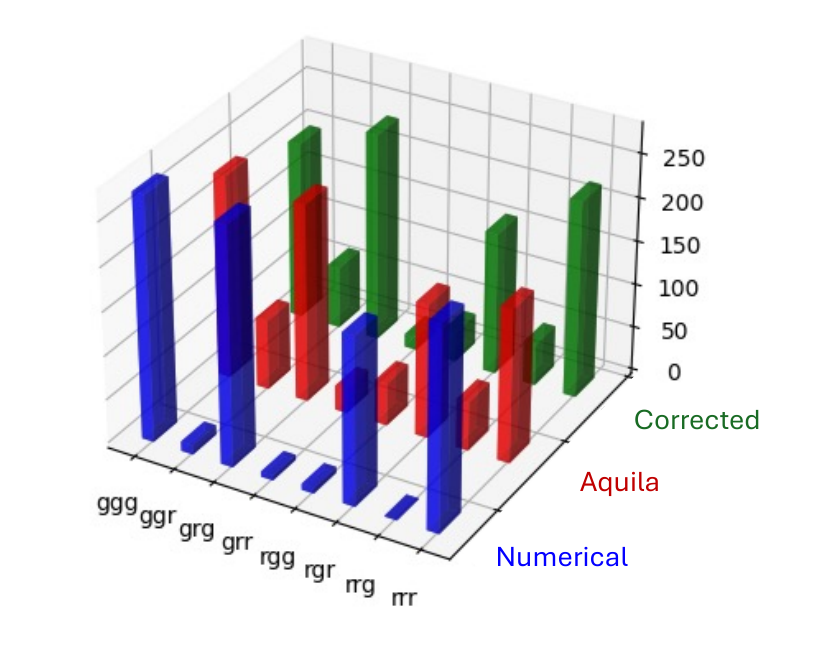}
\end{center}
\caption{ 
The same as Fig.~\ref{fig_sup_histogram_bell} but for the three atom chain graph state using the pulse sequence shown in Fig.~\ref{fig_cluster_state_circuits} with 1000 shots.  }
\label{fig_sup_histogram_3q}
\end{figure}

For uncorrelated classical measurement error, the single atom error generalizes to $N$-atom error.  For a single atom at $i=1$, $\bm{m}_{\text{meas}}$ has two entries: the number of counts for outcomes $\vert g \rangle$ and $\vert r \rangle$.  For one atom, the correction matrix has $2\times2$ elements and is given by:
\begin{equation}
\mathbb{T}_1\equiv
 \begin{bmatrix} 
    1 & \epsilon_M  \\
    0 & 1-\epsilon_M \\
    \end{bmatrix}.
\end{equation}
For $n$ independent measurements (one per atom) on atoms $i=1,...,N$, $\mathbb{T}$ has $2^n\times 2^n$ elements given by the tensor product of single atom matrices.  $\mathbb{T}$ is the matrix used to correct measurement bias discussed in the main text.  Figs.~\ref{fig_sup_histogram_bell} and \ref{fig_sup_histogram_3q} show improvements in example outcomes.

\section{Protocol for Post-Processing Measurements}
\label{sec_app_byproduct}

\begin{table}[t] 
\vspace{1cm}
\centering
\begin{tabular}{lcr}
\hline
Outcome  & Operation & Choice \\
\hline
10 & $\hat{I}$ & Discard \\ 
11 & $\hat{I} $ & Discard \\
01 & $\exp[-i\gamma \bm{n}\cdot\hat{\bm{\sigma}}/2]$ & Keep \\
00 & $\exp[-i\gamma\bm{n}\cdot\hat{\bm{\sigma}}/2]$ & Keep\\
\hline
\end{tabular}
\caption{Protocol for processing measurements in the first three atoms in a chain graph where the first atom is intentionally translated by $\Delta d$ (See Fig.~\ref{fig_logical_bloch}). The first column lists the four possible outcomes of measurements in the pseudospin eigenbasis of $\hat{\sigma}^x$ of the first two atoms.  The second column lists the unitary that impacts the third atom due to the projective measurements on atoms 1 and 2.  The third column shows what data to discard so that the pseudospin state of the third atom is rotated by an angle $\gamma$ on the Bloch sphere.} 
\label{table_prob} 
\vspace{0.1cm}
\end{table}

The graphs depicted in Figs.~\ref{fig_logical_bloch}a-c yield outputs that are the results of logic operations on input states after measurements in the pseudospin eigenbasis of $\hat{\sigma}^x$.  A feed-forward of the measurement results of each atom is needed to interpret the output(s).  The output of measurements for each atom is recorded and used offline in a byproduct unitary applied to the state(s) measured at the output(s).  These byproduct operators are derived for cluster states in Ref.~\cite{Raussendorf2003}.  For the identity operation in the chain  graph ($\Delta d=0$ in Fig.~\ref{fig_logical_bloch}a) the byproduct unitary is $\hat{U}_{\Sigma}=(\hat{\sigma}^z)^{s_1+s_3+..}(\hat{\sigma}^x)^{s_2+s_4+..}$, where $s_i=0,1$ are the measurement outcomes of the $i^{\text{th}}$ atom.  The following discusses the protocol defining $\hat{U}'_{\Sigma}$ in Eq.~\eqref{eq_rotated_output}.

To rotate the input state using distorted graphs with $\Delta d\neq0$ (Fig.~\ref{fig_logical_bloch}d), we must construct a new feed-forward protocol for $\hat{U}'_{\Sigma}$.  Consider three atoms where the first is a distance $d+\Delta d$ from the second but atoms 2 and 3 are separated by $d$ such that the interaction pulse time approximates a CZ gate between atoms 2 and 3.  Measuring atoms 1 and 2  in the eigenbasis of $\hat{\sigma}^x$ leads to four possible outcomes, listed in the first column of Table~\ref{table_prob}.  By starting with the eigenstate corresponding to the 3-atom perturbed graph state it is straightforward to show that a partial trace of the projective measurements performed on the first two atoms leads to a rotated output state at atom 3 only if outcomes 01 or 00 are obtained on the first two atoms.  Outcomes 10 and 11 have no impact.  The second column in Table~\ref{table_prob} lists the effective operation enacted on $\vert +_3\rangle$ by the measurement process.  The data shown in Fig.~\ref{fig_cluster_state_fidelity} only shows results where the first two atoms have outcomes 01 or 00.  This choice has the effect of rotating the state of the third atom for use as an input state to the remaining part of the graph where we then use the conventional cluster state byproduct operators, $(\hat{\sigma}^z)^{s_3+s_5+..}(\hat{\sigma}^x)^{s_4+s_6+..}$, for post processing.

\section{Trajectory Model for Teleportation Fidelity Scaling}
\label{sec_app_trajectory}

\begin{figure}[t]
\begin{center}
\includegraphics[width=0.46\textwidth]{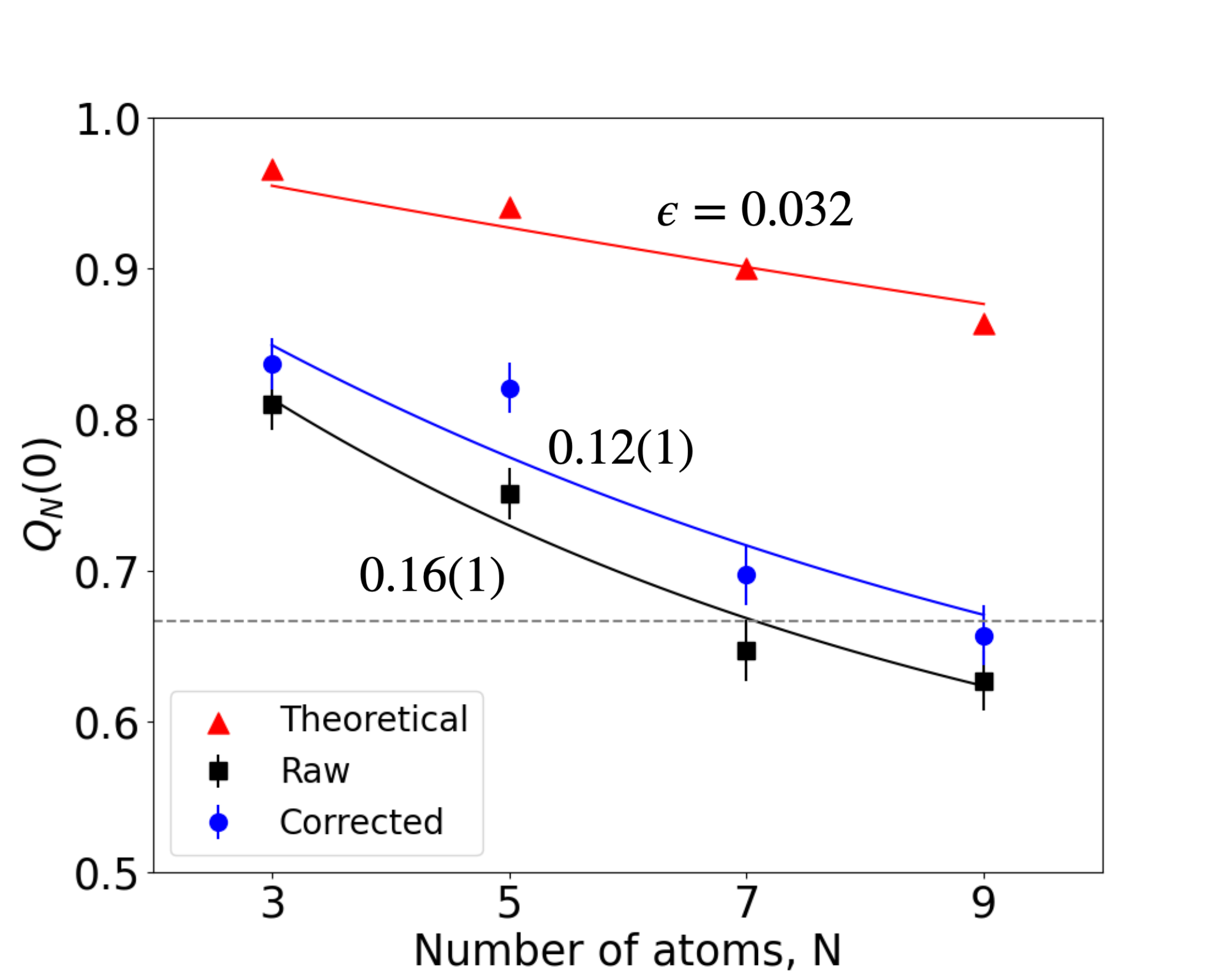}
\end{center}
\caption{ 
The symbols plot the same as Fig.~\ref{fig_scaling}.  The solid lines are single parameter fits with Eq.~\eqref{eq_fidelity_trajectory_method} to obtain $\epsilon$.  
}
\label{fig_app_scaling}
\end{figure}

In this section we consider an alternative fidelity model to fit teleportation scaling data discussed in Sec.~\ref{sec_aquila_computational_order}.  We rely on a quantum trajectory model \cite{CARLO2004} to study the impact of noise on the scaling of fidelity of the quantum state teleported along the chain graph.   We define the fidelity, $F_N$, as the overlap of the initial state with the output state.  As measurements move the initial state along the chain graph, noise perturbs the state.  Each step can be modeled as a move along a trajectory randomized by noise.  The resulting teleportation fidelity along an $N$ atom chain graph, assuming amplitude damping noise, is given by \cite{CARLO2004}:
\begin{align}
F_N=\frac{1}{2}\big(1+e^{-\epsilon N} \big), 
\label{eq_fidelity_trajectory_method}
\end{align}
where $\epsilon$ is the probability that the state of a single atom decays.  Note that for $\epsilon=\epsilon_L$, $F_N$ and $P_e(N/2)$ both become $1-\epsilon N/2 +\mathcal{O}(\epsilon^2)$ for low error probability.

Figure~\ref{fig_app_scaling} plots the same as Fig.~\ref{fig_scaling} but using a standard fit with Eq.~\eqref{eq_fidelity_trajectory_method}.  Here we see that the extracted error probability for the corrected data, $\epsilon=0.12(1)$, is near the value shown in Fig.~\ref{fig_scaling}, $\varepsilon_L=0.090(9)$.  The comparison in this section shows that exponential decay fidelities based on either dominate measurement errors or amplitude damping errors yield similar scaling parameterizations. 

\bibliography{references} 

\end{document}